%% file: ms.tex
\documentclass[ngerman,english,charter]{article}

\usepackage{arxiv}

\usepackage[utf8]{inputenc}
\usepackage{import}
\usepackage{float}
\usepackage{graphicx}
\usepackage{dsfont}
\usepackage{pgfplots}
\usepackage{caption}
\usepackage{subcaption}
\usepackage{amsmath}
\usepackage{amssymb}
\usepackage{hyperref}
\usepackage[bitstream-charter]{mathdesign}

\pgfplotsset{compat=1.14}

\title{Generating Steady-State Chain Fountains}

\date{} 					

\author{
  Johannes Mayet\\
  Chair of Applied Mechanics\\
  Department of Mechanical Engineering\\
  Technical University of Munich\\
  \texttt{johannesmayet@tum.de} \\
   \And
  Friedrich Pfeiffer\thanks{\url{www.amm.mw.tum.de}}\\
  Chair of Applied Mechanics\\
  Department of Mechanical Engineering\\
  Technical University of Munich\\
  \texttt{pfeiffer@tum.de} \\
}

\providecommand{\sDelta}{{\scriptstyle{\Delta}}}
\providecommand{\vec}[1]{\boldsymbol{#1}}
\providecommand{\bmat}[1]{\begin{bmatrix} #1 \end{bmatrix}}

\begin{document}

\maketitle

\begin{abstract}

In recent years the chain fountain became prominent for its counter-intuitive fascinating physical behavior. Most widely known is the experiment in which a long chain leaves an elevated beaker like a fountain and falls to the ground under the influence of gravity. The observed chain fountain was precisely described and predicted by an inverted catenary in several publications. The underlying assumptions are a stationary fountain and the knowledge of the boundary conditions, the ground and beaker reaction forces. In contrast to determining the steady-state chain fountain shape, it turns out that the main difficulty lies in predicting the reaction forces. A consistent and complete physical explanation model is currently not available. In order to give a reasonable explanation for the reaction forces an illustrative mechanical system for generating steady-state chain fountain is proposed in this work. The model allows to generate all physical possible chain fountains by adjusting a pulley arrangement. The simplifications incorporated make the phenomenon accessible to undergraduate students.
\end{abstract}

\section{Introduction}
\input{section/introduction.tex}
\section{Equations of Motion}
\input{section/belt_approximation.tex}
\section{Pick-Up and Pull-Down Forces}
\input{section/pickup_momentum_approach_v01.tex}

\section{Generating Steady-State Chain Fountains}
\input{section/apparatus.tex}

\section{Conclusion}
As a summary, the derivation of the chain fountain with simplified spherical joint resistance torque is presented. It was shown that identical fountains can be generated with completely different force characteristics.  The key to this is the insight that already the kinematic parameters $l$, $h$ and $H$ are completely defining the fountain. Since the free chain is just cut out of the overall system, the kinetic parameters cannot be uniquely determined with such a subsystem consideration. As a consequence, only value ranges can be specified. The core of this work, however, is to find a simple explanation for the necessary cutting forces. The presented substitute system for the unfolding and folding process allows the whole range of values for the reaction forces to be covered. Even if it is not possible to entirely explain the chain fountain in the cited videos with the proposed mechanism, it has been shown that a steady-state chain fountain can be generated without additional argumentation such as traveling waves, dissipation effects, shock propagation, transverse motions or ground kicks. 
\bibliographystyle{plain}
\bibliography{ms}

\end{document}

%% file: section/introduction.tex
\begin{figure}[!h]
\centering
\begin{subfigure}[b]{0.24\textwidth}
\includegraphics[width=1\linewidth]{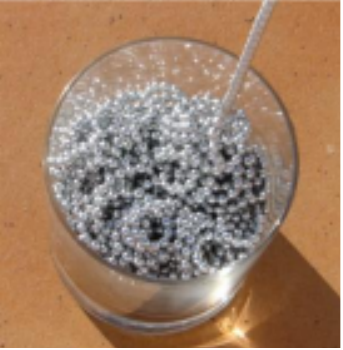}
\subcaption{Initial configuration.}
\label{fig:initsetting_chaotic}
\end{subfigure}
\begin{subfigure}[b]{0.24\textwidth}
\includegraphics[width=1\linewidth]{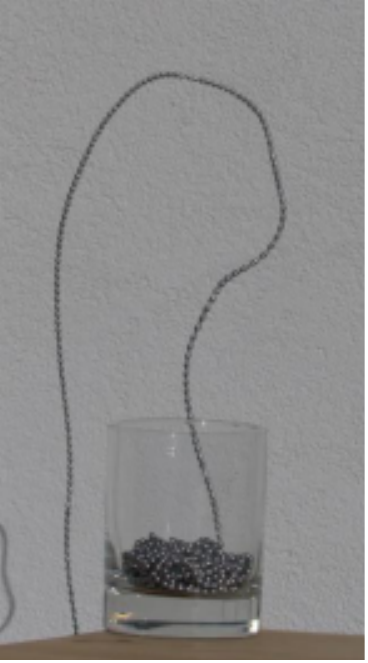}
\subcaption{Fountain of (a).}
\label{fig:fountain_chaotic}
\end{subfigure}
\begin{subfigure}[b]{0.24\textwidth}
\includegraphics[width=1\linewidth]{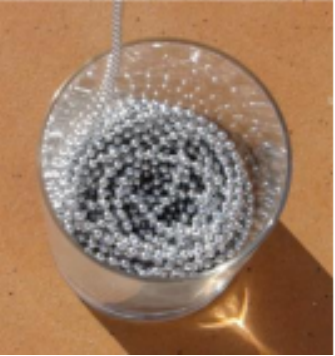}
\subcaption{Initial configuration.}
\label{fig:initsetting_ordered}
\end{subfigure}
\begin{subfigure}[b]{0.24\textwidth}
\includegraphics[width=1\linewidth]{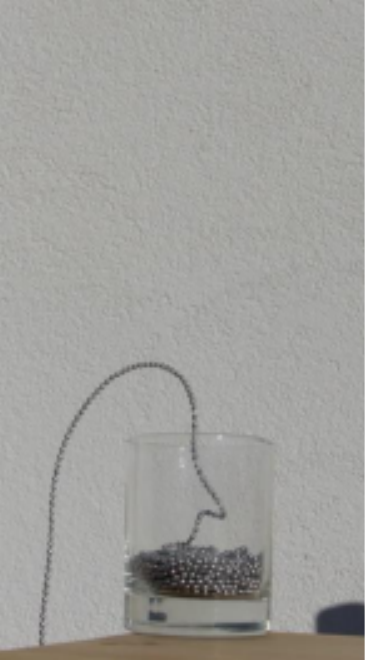}
\subcaption{Fountain of (c).}
\label{fig:fountain_ordered}
\end{subfigure}
\caption{The random initial setting shown in (a) yields a much higher fountain depicted in (b) compared to the initial setting (c) and corresponding fountain (d) when falling from the same height (pictures are taken from~\cite{martins2016initial}).}
\label{fig:martins}
\end{figure}

Since 2013 when Steve Mould~\cite{mould2013selfsiphoning,mould2014tedx} presented its ``self-siphoning beads'' in a video, the chain fountain mystery has reached the public attention. Beginning with the 25th International Young Physicists' Tournament in 2012 the problem has been addressed by many scientists nowadays. In the chain fountain experiment one observes how a long chain leaves an elevated beaker like a fountain and falls to ground under the influence of gravity (cf. Fig.~\ref{fig:martins}). 
From a scientific point of view the underlying mechanical problem can be split into two separate problems. First, one has to determine an appropriate chain model, e.g. inextensible string or connected rigid bars, with which it is possible to represent chain properties like dissipation and bending stiffness in a quantitative manner in order to address specific phenomena, e.g. traveling waves. Second, one faces the challenge to describe the (averaged) dynamics of folding and unfolding chains because it needs a mechanism to accelerate the chain links from their resting state to constant speed and vice versa. These problems are very well known since a long time, see for example Routh~\cite{routh1860} or Hamel~\cite{hamel1949theoretische}, who consider continuous chain models. Certainly, publications concerning the paradoxical phenomenon that the free end of a vertically hanging folded chain is falling faster than a free-falling body under gravitational acceleration have also to be considered. This topic is addressed by Schagerl~\cite{schagerl1997paradoxfallingchain} and Steiner~\cite{steiner1995foldedinextensible} in experiments and numerical simulations, for example. In reference~\cite{hamm2010} it is shown that the geometry of the folding at the bottom is crucial and has to be accounted for in the dynamics of the chain. The falling chain experiments indicate that a ball chain with a minimal radius of curvature is more accelerated in contrast to classical loop chains with no well-defined minimum radius of curvature (cf.~\cite{hamm2010}). Grewal~\cite{grewal} suggests different mechanical designs to increase the phenomenon of a down-pulling falling chain. Willerding~\cite{will2014} considers space problems connected with chains and ropes, but also the falling U-shape chain and the chain fountain. \newline
At the International Young Physicists Tournament (\textit{IYPT}) in 2012 several interesting measurements gave attention to unfolding mechanisms of the chain. Hledik~\cite{hledik2012iypt} and Santiago~\cite{santiago2012iypt} show results with initial chain arrangements which are well-ordered (see Fig.~\ref{fig:initsetting_ordered}) or chaotic (see Fig.~\ref{fig:initsetting_chaotic}) yielding completely different fountains, as illustrated in Fig.~\ref{fig:fountain_ordered} and Fig.~\ref{fig:fountain_chaotic}. This result has been recently confirmed by Martins~\cite{martins2016initial}. These observations are conceptually in agreement with the unfolding process as dissipationless extraction of chains on a flat table, cf. reference~\cite{Yokoyama2018ReexaminingTC,hanna2012slackdynamics}. Flekkøy, Moura and Måløy~\cite{flekkoy2018fluctuations} study based on simulations different types of chains as well as surface conditions of the beaker. On the one hand, the importance of a rough beaker bottom is emphasized since it causes a different unfolding behavior of the chain. On the other hand, the relevance of different chain types, which are modeled by different internal stiffness interactions, is pointed out. Virga~\cite{virga2014shocks}, however, uses a continuous chain model and applies "dissipative shocks" at the beaker and floor by arguing that the pick-up and put-down processes take place in a very short time. Interestingly collisions are a key concept in Grewal's~\cite{grewal} designs increasing the falling chain phenomenon. Although shock propagation and internal stiffness interactions are completely different approaches, both theories yield convincing results with respect to characteristic fountain parameters.\newline
Biggins and Warner~\cite{biggins2014} derive an inverted caternary fountain by using a simple inextensible string model. The inverted caternary is in good agreement with their experimental results and those of Pantalenone's~\cite{pantaleone2017quantitative} quantitative study. The inverted caternary fountain, however, requires reaction forces on the starting point and endpoint of the fountain. The associated reaction forces are interpreted in reference~\cite{biggins2014} by collisions of each chain element with the support when lifted up. Despite the fact that the influence of different initial chain arrangements can not be explained reasonable, this approach is quite controversial due to high-speed camera videos showing that the chain elements can leave their support in a horizontal direction for a long time without affecting the overall chain fountain.\newline
In contrast to the above-mentioned publications the aim of this work is to present a simple principle model of the chain fountain for which it is possible to derive the steady-state behavior by fundamental mechanical laws. Speculative forces arising from the table and floor will not be required since the beaker and the floor interactions are replaced by a pulley arrangement allowing for a continuous acceleration and deceleration of the chain elements. As a consequence, it is not necessary to deal with shock propagation and traveling waves. However, the enormous advantage of this principle model is also the biggest disadvantage, since the complex system behavior can only be approximated. No conclusions can be drawn about the real folding and unfolding mechanism.\newline
The paper is organized as follows: In the first part the inverted caternary is derived by assuming apriori known beaker and floor interaction forces yielding the free flight dynamics of the chain. The following section discusses the principle model for folding and unfolding the chain at the end points, which allows the explanation of reaction forces within their limits. Finally, the overall principle model of the chain model is presented and the underlying equations for a steady-state fountain are given. 

%% file: section/belt_approximation.tex
\providecommand{\parDs}{\frac{\partial}{\partial s}}
\providecommand{\parDxDs}{\frac{\partial \vec{x} }{\partial s}}
In this section the equations of motion for a steady-state chain from Pfeiffer~\cite{pfeiffer2017chainfountain} are recapitulated. In contrast to the equations of motion for thin strings the equations are extended by additional rotational stiffness and damping effects. However, it is assumed that the chain is inextensible ($||\vec{x}'||=1$ where $()'=\partial/\partial s $ is the partial derivative w.r.t. the arc-length $s$ of the fountain), the density is constant and more importantly the velocity is constant ($||\dot{\vec{x}}||=v$). A useful choice for the representation of the velocity is then given by $\dot{\vec{x}}=v\begin{bmatrix} \cos(\alpha) & \sin(\alpha) \end{bmatrix}^T$ since the assumptions $||\dot{\vec{x}}||=v$ and $||\vec{x}'||=1$ are satisfied from the beginning. 
\begin{figure}[!h]
\centering
\begin{subfigure}[b]{0.35\textwidth}
\centering
\includegraphics[width=0.5\linewidth]{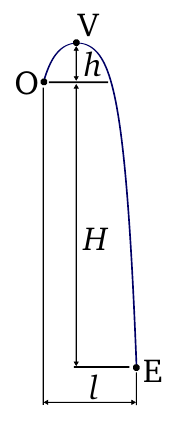} 
\subcaption{Chain fountain parameters.}
\end{subfigure}
\begin{subfigure}[b]{0.5\textwidth}
\centering
\includegraphics[width=1\linewidth]{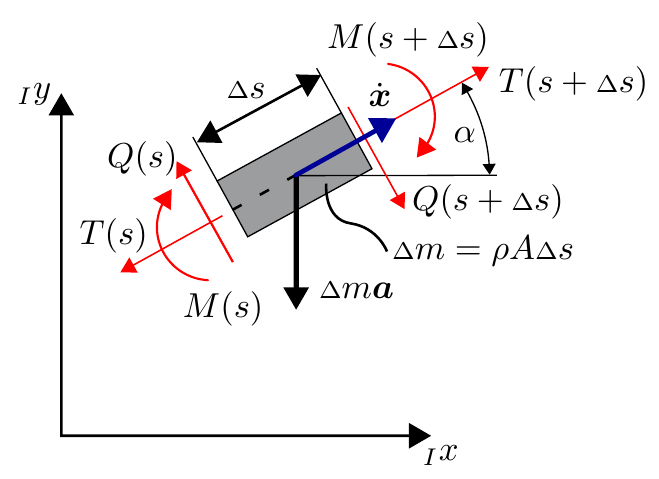}
\subcaption{Infinitesimal belt segment and cutting forces.}
\label{fig:chain_element_intro}
\end{subfigure}
\caption{The chain fountain is described by its overshoot height $h$, falling down height $H$ and horizontal distance $l$ between the beaker and the approximated landing area on the floor (left). Internal forces in tangential and normal direction as well as a bending torque act on a gravitational attracted infinitesimal belt segment of mass $\sDelta m$ and length $\sDelta s$ (right).}
\end{figure}
The linear momentum for an infinitesimal chain element yields
\begin{align*}
\ddot{\vec{x}}(s,t)  \lambda(s) \sDelta s &= - \vec{T}(s,t) + \vec{T}(s+\sDelta s,t) - \vec{Q}(s,t) + \vec{Q}(s+\sDelta s,t) +  \vec{a}(s,t) \lambda(s) \sDelta s\,,
\end{align*}
where $\lambda(s)$ is the linear mass density $\rho(s)$ times cross-sectional area $A$, $\vec{T}$ and $\vec{Q}$ are vectors of internal forces and $\vec{a}(s,t)$ is the vector of gravitational acceleration acting on the segment with length $\sDelta s$, see Fig.~\ref{fig:chain_element_intro}. The curve of the chain is described by $\vec{x}=\begin{bmatrix}x_p & y_p\end{bmatrix}^T$ with the choice $x_p = y_p = 0$ at the vertex. By approaching the limit $\sDelta s \to 0$ one obtains a continuous partial differential equation:
\begin{align}
 \lambda(s) \ddot{\vec{x}}(s,t) = \frac{\partial}{\partial s}\vec{T}(s,t) + \frac{\partial}{\partial s}\vec{Q}(s,t) + \lambda(s) \vec{a}(s,t)\,.
\end{align}
 The internal cutting forces are chosen such that $\vec{T} = T(s,t) \begin{bmatrix}\cos(\alpha) & \sin(\alpha)\end{bmatrix}^T$ acts in direction of the tangent vector and $\vec{Q} = Q(s,t) \begin{bmatrix}\sin(\alpha) & - \cos(\alpha)\end{bmatrix}^T$ acts in direction of the normal vector. Applying the steady-state assumptions ($\dot{s}=v$) the equations of motion can be written as 
\begin{align}
 \alpha'\left(\bar{T}-1\right)+ \bar{Q}' &= k \cos\alpha && \text{and} && \bar{T}'- \alpha'\bar{Q} = k \sin\alpha \,,
\end{align}
where the scaled internal forces $\bar{T}=T/(\lambda v^2)$ and $\bar{Q}=Q/(\lambda v^2)$, and the constant $k=g/v^2$ accounting for gravitational forces ($\vec{a}=[0\quad-g]^T$) have been introduced. Usually the internal forces $\bar{Q}$ are neglected for materials like ropes and strings since one assumes zero bending stiffness for such materials. In the case of the considered chain, which consists basically of small rigid rods connected by spherical joints, normal forces $\bar{Q}$ have actually to be considered even if the chain elements are assumed to be (infinitesimal) small. If a rotational spring (coefficient $c$) and damper (coefficient $d$) resistance of the spherical joints are introduced, then the angular momentum equation with mass moment of inertia $J \lambda\sDelta s$ is 
\begin{align*}
J \lambda\sDelta s \ddot{\alpha}(s,t) = & - \frac{1}{2} \sDelta s\left(Q(s + \sDelta s ,t) + Q(s ,t) \right) - c \left(\alpha(s + \sDelta s ,t)-\alpha(s ,t)\right)- d \left(\dot{\alpha}(s + \sDelta s ,t)-\dot{\alpha}(s ,t)\right)\,,
\end{align*}
and by approaching the limit $\sDelta s \to 0$ one obtains $\bar{Q} = - J \alpha'' - c/(\lambda v^2)\alpha' - d/(\lambda v)\alpha''$. Further, if a spherical joint resistance, which linearly depends on the tension along the curve, is assumed, one may postulate a mathematical motivated ansatz $\bar{Q} = - \left(\varepsilon_0 + \varepsilon_1 \bar{T}\right) \alpha' - \left(\varepsilon_2 + \varepsilon_3 \bar{T} \right) \alpha''$ with all $\varepsilon_i\in\mathbb{R}^+\ll1$ in order to apply perturbation methods. In this work, however, the influence of the parameters $\varepsilon_i$ are only addressed numerically from a phenomenological point, because experiments that reveal physical parameters are not available at the present time. In order to understand the underlying physical relationships of the chain fountain, the solution curves for all $\varepsilon_i=0$ are examined in the next section in detail.  
\subsection{Inverted Caternary}
\begin{figure}[!h]
\centering
\includegraphics[width=0.8\linewidth]{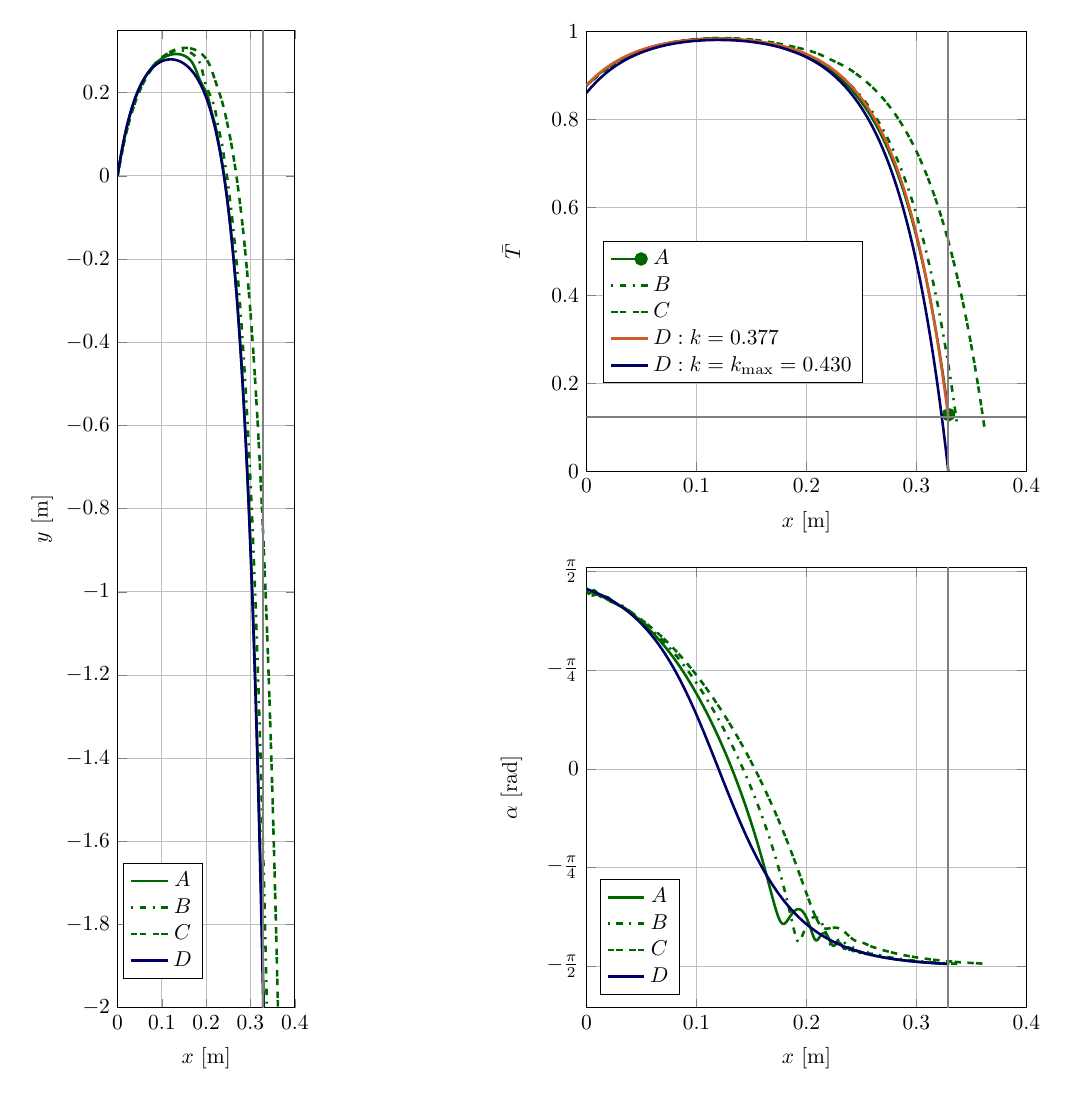}
\caption{Example for results obtained for a chain fountain using the belt approximation: left side the fountain, right side top the scaled internal chain tension, and right side bottom the fountain angle $\alpha$ for different parameters. Equal parameters: $H=2\,[m]$, $\xi=h/H=0.14$, $\kappa=l/H=0.165$, $Hq=44.4$, $\varepsilon_1=\varepsilon_3=0$.
Fountain A: $\varepsilon_1=1e-5,\,\varepsilon_2=2e-4$ and $k=0.377$. 
Fountain B: $\varepsilon_1=2e-5,\,\varepsilon_2=4e-4$ and $k=0.377$.
Fountain C: $\varepsilon_1=2e-5,\,\varepsilon_2=8e-4$ and $k=0.377$. 
Fountain D: $\varepsilon_1=\varepsilon_2=0$ and $k=0.377$ (blue) or  $k=0.430$ (orange).}
\label{fig:schematic_chain_fountain}
\end{figure}

In the case that all $\varepsilon_i$ are equal zero (cf. Biggins~\cite{biggins2014}), the resulting differential equations $\alpha'\left(\bar{T}-1\right)= k \cos\alpha$ and $\bar{T}' = k \sin\alpha$ lead directly to the solution for $\bar{T}=1-k/(q \cos\alpha)$ in terms of $\alpha(s)$ and one remaining differential equation $\alpha'= q \cos^2\alpha$ with $q=k\left[(1-\bar{T}_E)\cos\alpha_E\right]^{-1}>0$. The constant parameter $q$ takes into account that the tension at endpoint, given by $\alpha(s=s_E)=\alpha_E\in[0,\pi/2]$, is equal to $\bar{T}_E\in[0,1[$. Although determining a solution for $\alpha=\alpha(s)$ means no serious difficulties, it is advantageous to determine a solution for $y_p = y_p(\alpha)$ using $y'_p=\sin\alpha$ in order to establish the relation to the geometrical dimensions of the fountain:
\begin{align}
 \mathrm{d}y_p &= \sin\alpha\, \mathrm{d}s =  \sin\alpha \, \frac{\mathrm{d}s}{\mathrm{d}\alpha}\,\mathrm{d}\alpha && \rightarrow && \cos\alpha = 1/(1 - q y_p)
\end{align}
Using $\mathrm{d}y_p = \tan\alpha\, \mathrm{d}x_p$ 
we obtain the solution curve as function graph 
\begin{align}
 q y_p &= 1-\cosh\left(q x_p\right)\,. \label{equ:func_graph}
\end{align}
\newcommand{\Tmax}{\bar{T}_p}
This inverted caternary coincides with the results of Biggins~\cite{biggins2014}. This solution representation stands for its simplicity. Further, a severe conclusion can be drawn: As it is possible to scale lengths by $q$, e.g. $\bar{y}_p = q y_p$ and $\bar{x}_p = q x_p$, yielding a parameter independent solution curve, it is obvious that one can not state an additional mechanical law for the free flight dynamics which can be used to determine the unknown parameter $q$ respectively the unknown reaction forces.
Note that this actually means that the constants $q$ and $k$ arising in the cutting force $\bar{T}  = 1 - k/q +  k y_p < 1$ are either already determined by kinematic quantities or must be identified by an overall system approach. In our derivation the forces $T_0$ and $T_E$ occur at the endpoints and are assumed to be constant due to the steady-state assumption. But questions on how these cutting forces really originate can not be answered since an overall system model would be required. \newline  
Employing again the function graph representation given in Eq.~\eqref{equ:func_graph} we derive the relationship $\cos\alpha = \left(1 - q y_p\right)^{-1} $ which is used to obtain an alternative formulation for the endpoint angles  
by taking the beaker location at height $y_p=-h$ and the floor height $y_p=-h-H$ into account, see Fig.~\ref{fig:chain_element_intro}:
\begin{align}
q l = \text{acosh}\left(1 + q (h+H)\right) + \text{acosh}\left(1 + q h \right)
\label{equ:kinemat_constraint}
\end{align}
\begin{figure}
\centering
\includegraphics[width=0.85\linewidth]{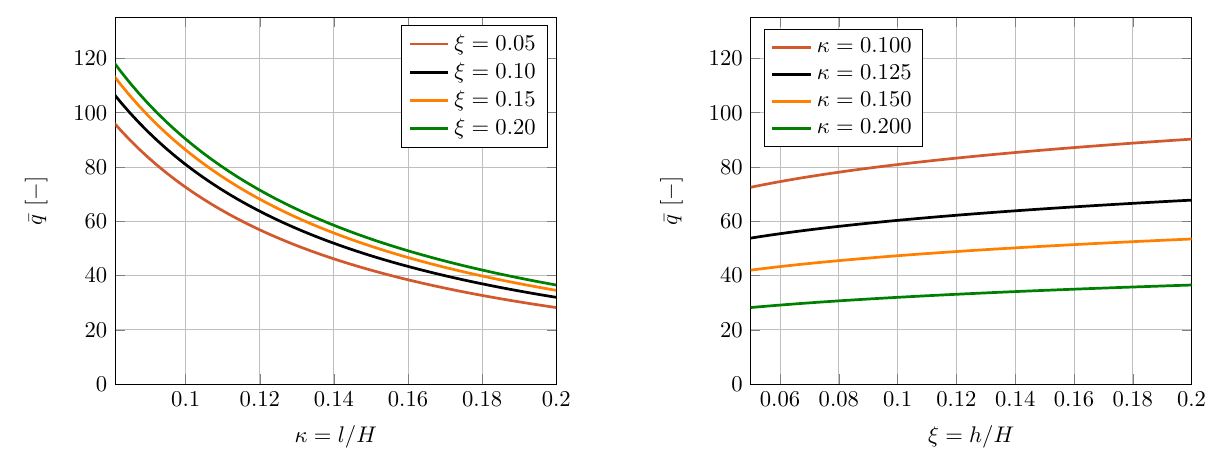}
\caption{Numerical solution of $\bar{q}=Hq$ using Eq.~\eqref{equ:kinemat_constraint} for given values of $\xi=h/H$ and $\kappa=l/H$.}
\label{fig:sol}
\end{figure}
Consequently, the parameter $q$ is implicitly given by Eq.~\eqref{equ:kinemat_constraint} and therefore 
the fountain is completely specified by three points (beaker, floor and vertex), respectively the parameters $l$, $h$ and $H$. In Fig.~\ref{fig:sol} numerical solutions of Eq.~\eqref{equ:kinemat_constraint} for $\bar{q}=Hq$ are depicted. It is important to note that the shape of the fountain can not be altered in the case of the proposed mechanical model.  A physical interpretation of the parameter $q$ is revealed by evaluating the signed curvature $(x'y''-x''y')/({x'}^2 + {y'}^2 )^{3/2} = \alpha' = q \cos^2\alpha$ at the vertex with $\alpha=0$. This result is by far not intuitive since it might be expected for example that fountains can be realized having different shapes due to different traveling velocities.\newline
Since $q$ is already defined by kinematic relationships the remaining unknown parameter $k=g/v^2$ has to be determined by measuring the average velocity of the chain elements. 
Although measuring the average velocity does not present any particular difficulties, it useful to restrict the possible parameter range for $k$ by assuming tensile forces:
\begin{subequations}
\begin{align}
 \bar{T}_0 & = 1 - \frac{k}{q}  - k h\geq0\\
 \bar{T}_E & = 1 - \frac{k}{q}  - k (h+H)\geq0 && \rightarrow && k\leq \frac{q}{1+(h + H)q} < \frac{1}{h + H}
 \label{equ:k_upper_bound}
\end{align}
\end{subequations}
Therefore, it can be concluded using the above inequality for $k$ that the average velocity $v$ has to be greater than $v_\text{min}=\sqrt{g(h+H)}$. The velocity $v_\text{min}$ is equal to the vertical velocity of a rigid body at ground when dropped at height $h+H$ with zero initial vertical velocity. In the case of $\bar{T}_E\approx0$ and $q\gg1$ the average velocity $v$ and the minimal velocity $v_\text{min}$ are indeed approximately equal. In order to obtain a lower bound for the parameter $k$ it is assumed that conservation of energy for the free flight chain segment holds, which yields $v_\text{max}=\sqrt{2gH}$. The resulting lower bound $k>1/(2H)$ and as a result $qH > 1/(1-\xi) > 0$ with $\xi=h/H\in[0,1[$ are very rough estimates at best. But without measuring the average velocity $v$ it is now possible to restrict the end-point forces:
\begin{subequations}
\begin{align}
0\,\leq\quad&\bar{T}_E\quad< \, 1-\frac{1}{2}\left(\frac{1}{qH}+1+\xi\right)\, < \, \frac{1}{2}\\
\frac{1}{2}\, < \, \frac{qH}{1+qH(1+\xi)}\, \leq \quad&\bar{T}_0 \quad <\, 1-\frac{k}{q}\, < \, 1-\frac{1}{2qH} \, <\, \frac{1}{2}(1+\xi) \,< \,1
\end{align}
\label{equ:range_forces}
\end{subequations}
As a consequence, it can be stated that $\bar{T}_E\in S_E \subseteq[0,1/2[$ and $\bar{T}_0 \in S_0 \subseteq\,]1/2,1[$ has to be satisfied for all chain fountains using a continuous string model. However, as previously explained it is not possible to identify unique values for $\bar{T}_E$ and $\bar{T}_0$ without measuring the average velocity $v$ of the chain. 
In Fig.~\ref{fig:schematic_chain_fountain} an example for a chain fountain (fountain \textbf{D})  is depicted with two different values for $k$ ($v=4.8\,[m/s]$ and $v=5.1\,[m/s]$) as it may be observed in the experiment. As expected, the fountain cannot be distinguished since the chain parameter $l$, $h$ and $H$ are the same in both cases. However, in the case of $k_\text{max}=q/\left(1+ (1 + \xi)qH\right)$ the ground reaction force $\bar{T}_E$ is equal zero in contrast to the case where the chain moves $0.3\,[m/s]$ slower. As illustrated in Fig.~\ref{fig:schematic_chain_fountain}, it must be kept in mind that these statements are only permissible under the assumption of a string model. The additional internal force $\bar{Q}$ which is motivated by the spherical joint resistance and mass moment of inertia gives rise to chain fountains with possibly higher overshoot heights $h$. Different heights do not necessarily cause significant differences in horizontal displacement $l$ as depicted by fountains \textbf{A} and \textbf{B}. Conversely, different horizontal displacements can be generated whereby maintaining the overshoot height.
\newline
As an intermediate conclusion it can be said that the presented chain model approximates the complex dynamical behaviour reasonably well. It is essentially useful when ruling out non-physical parameters or reaction forces, but discrepancies will inevitably remain.

%% file: section/pickup_momentum_approach_v01.tex
\begin{figure}[!h]
\centering
\includegraphics[width=0.9\linewidth]{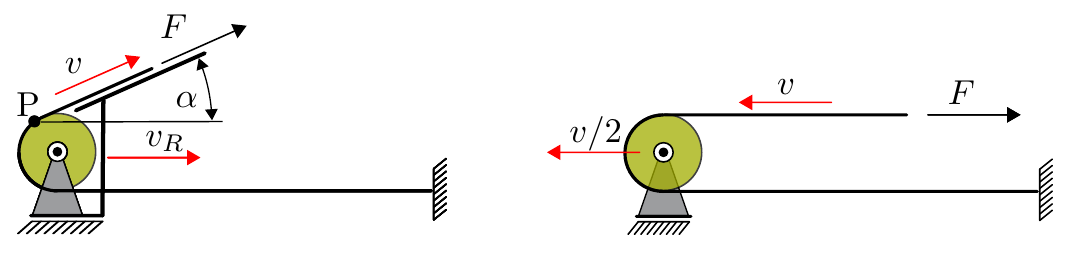}
\caption{Schematic steady-state unfolding and folding process: The limited curvature of the chain is taken into account by the rollers. Left: As the upper part of the chain travels with velocity $v$ the mass increase is $dm = \lambda v_R \mathrm{d}t$ with $v_R=v/(1+\cos(\alpha))$. As a consequence, the pulling force $F$ has to be $F=\lambda v^2/(1+\cos(\alpha))$. Right: As the upper part of the chain travels with velocity $-v$ the mass decrease is $dm = \frac{1}{2} \lambda v \mathrm{d}t$. As a result the pull-down force $F$ is given by $F=\frac{1}{2} \lambda v^2$. 
}
\label{fig:schematic_pick_up}
\end{figure}

The question which naturally arises is how one can obtain the specified ranges for the forces given in Eq.~\eqref{equ:range_forces} by picking up chain elements. Commonly one is tempted to average the dynamical behavior by applying the law of conservation of momentum for a particle, which is at rest and afterwards it travels with velocity $v$ within an infinitesimal short time, leading to 
\begin{subequations}
\begin{align}
 \int_{t^-}^{t^+}T_0\mathrm{dt}&=\Delta m (v(t^+)-v(t^-)) && \rightarrow && \bar{T}_0 = 1\,, 
\end{align}
\end{subequations}
which is not in agreement with the results of the belt approximation because $\bar{T}_0$ has to be smaller than the maximal force $T_\text{max}=1-k/q<1$. Since the chain is treated as a chain of particles this approach approximates the pick-up process well if the chain is made of single elements connected by strings. In our case, however, the chain has the distinctive property of a minimal radius of curvature with the result that the chain is unfolded rather than that individual chain particles are accelerated by an impact. Therefore, we follow in this work a simple contemplation to justify that the scaled force $\bar{T}_0$ does not necessarily has to be close to $1$. As illustrated in Fig.~\ref{fig:schematic_pick_up} the pick-up process is described by chain with zero bending stiffness guided by a massless roller. As depicted on the left side in Fig.~\ref{fig:schematic_pick_up} a pulling force $F=\lambda v^2/(1+\cos\alpha)$ is required to reach steady-state. This result is obtained by noting that the velocity of the point \textsf{P} is given by $\dot{\vec{x}}_P=\bmat{v_R & 0}^T- r\dot{\alpha}\bmat{\cos\alpha & \sin\alpha}^T$ with roller radius $r$. The angular velocity is given by $\dot{\alpha}=-v_R/r$, because the lower part of the chain is not moving. The velocity of point \textsf{P} along the inclined direction is then $\bmat{\cos\alpha & \sin\alpha}\dot{\vec{x}}_P = v_R (1+\cos\alpha)$. Consequently, we have $v_R=v/(1+\cos\alpha)$, which is the translational velocity of the roller and therefore the mass increase velocity. Conservation of momentum then requires a pulling force $F=\lambda v^2/(1+\cos\alpha)$. In a first limit case consideration, it can be concluded that the true scaled force probably originates from a mechanism that is something between this simplified unfolding mechanism with $\alpha\in[0,\pi/2[$ and a pure impulse that is equivalent to $\alpha=\pi/2$.
On the other hand, it is also possible to estimate the pull-down forces acting on the chain with the same basic consideration. Assuming that the pull-down process is an average of chain elements hitting directly the table with $F=0$ and a folding mechanism, it is expected that the resulting forces lie within the desired value range. The unfolding respectively folding of the chain over the roller, which ensures the minimum radius of curvature, is therefore useful to explain how the defined value range given by Eq.~\eqref{equ:range_forces} can be covered. Of course, the connection to other investigations is present, e.g. the phenomenon of the U-shape or the ``sucking'' chain~\cite{grewal,schagerl1997paradoxfallingchain, steiner1995foldedinextensible}. These observations permit now the construction of an apparatus with which steady-state chain fountains can be generated.

%% file: section/apparatus.tex
\begin{figure}[!htb]
\centering
\includegraphics[width=1\linewidth]{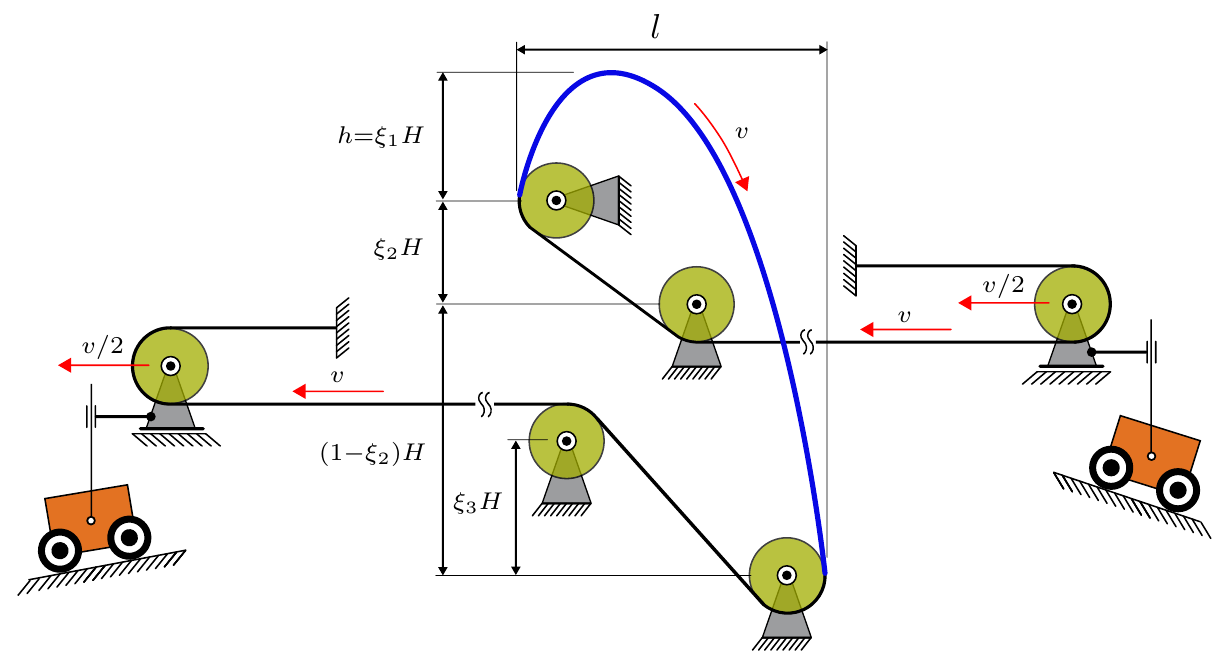}
\caption{An arrangement of pulleys to decelerate (left) and accelerate (upper right) the chain elements from rest to constant velocity $v$ and vice versa in order to generate a steady-state chain fountain (blue). The additional carts on an inclined plane allow to manipulate the tension in order to get different values than $\bar{T}=1/2$, see also Fig.~\ref{fig:schematic_pick_up}. Positioning the "bottom'' by $\xi_3H$ above the chain endpoint and the ``beaker'' $\xi_2H$ under the starting point of the chain gives the opportunity to make further adjustments. It should be emphasized that the rollers are displayed over-proportionally.}
\label{fig:pulley_experiment}
\end{figure}
\begin{figure}[!htb]
\centering
\includegraphics[width=1\linewidth]{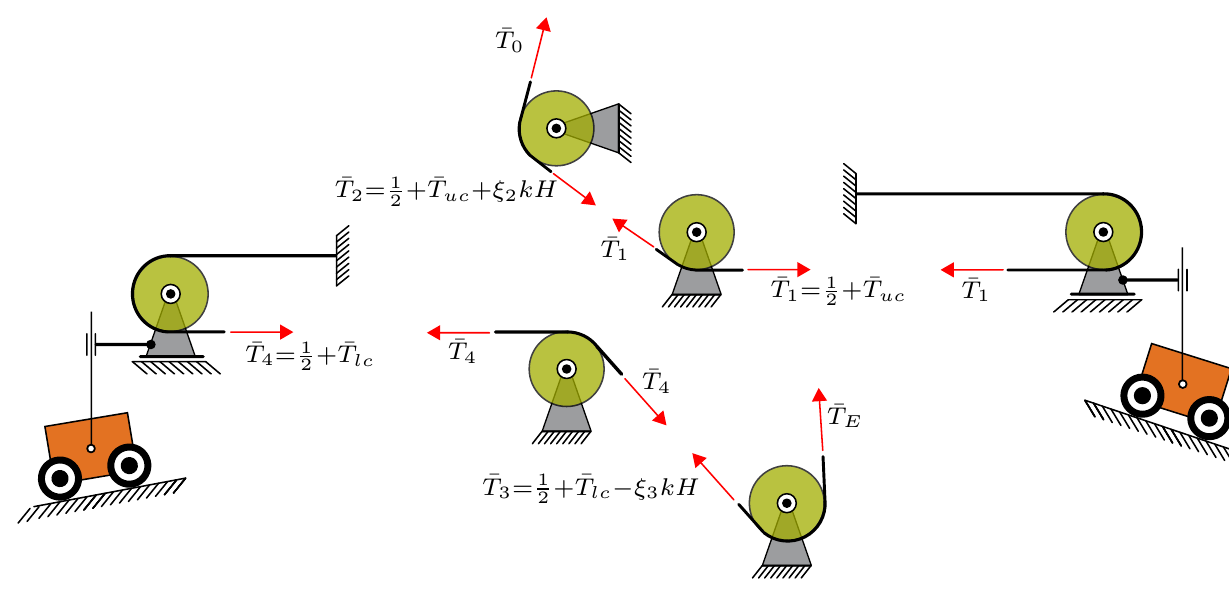}
\caption{At crucial transition points the resulting reaction forces/tensions are illustrated. Additionally to the folding, respectively unfolding mechanism from which one gets $\bar{T}=1/2$, the carts are accounted for by $\bar{T}_{lc}$ and $\bar{T}_{uc}$, which are scaled quantities. The ``floor'' was raised in relation to the chain endpoint in order to be able to generate values for $\bar{T}_E$ smaller than $1/2$. The ``beaker'' was lowered in relation to the starting point of the fountain even though the cart has the same influence. In both cases, the effect on the tension due to gravity was considered. The influence of the rollers on the balance of forces was neglected for the sake of simplicity.}
\label{fig:pulley_experiment_cutting}
\end{figure}

In order to be able to generate steady-state chain fountains we use the above presented concept of guiding pulleys. In Fig.~\ref{fig:pulley_experiment} an example for an arrangement of the pulleys is illustrated. Due to the pulleys it is possible to continuously accelerate and decelerate the chain elements. Since the radius of the roller is irrelevant for the pulling force, it is assumed that the radius is sufficiently small and the influence on the force balance can be neglected for the sake of simplicity. Furthermore, the principle model should be considered idealized in terms of non-conservative influences such as dissipation and friction. As depicted in Fig.~\ref{fig:pulley_experiment_cutting} it is possible to obtain reaction forces $\bar{T}_E=\frac{1}{2} + \bar{T}_{lc} - \xi_3kH$ and $\bar{T}_0=\frac{1}{2} + \bar{T}_{uc} + \xi_2kH$ where $\bar{T}_{lc}>0$ and $\bar{T}_{uc}>0$ are the scaled gravitational forces of the upper ("uc``) and lower cart (''lc``) on an inclined track. In order to reach steady-state the equation $\bar{T}_0-\bar{T}_E=kH$ has to be satisfied. This can be achieved by setting $\bar{T}_{uc} = \bar{T}_{lc} + kH\left(1-\xi_2-\xi_3\right)$ and as a result 
$\bar{T}_0=\frac{1}{2} + \bar{T}_{lc} + kH\left(1-\xi_3\right) $. Certainly, this requirement is equivalent to conservation of energy which is consistent with our assumptions.
The enormous advantage of this approach is that the chain fountain can be determined in a two-step procedure. First the fountain kinematics are defined and the associated forces of the free chain are calculated. In the next step, the remaining heights and the cart-masses would be chosen in such a way that the necessary forces are achieved. It should be kept in mind that it is theoretically possible to generate all relevant reaction forces if the free flight dynamics of the chain are captured accordingly with the proposed simple string model. Taking into account the already derived value ranges of the reaction forces one obtains  $\xi_3 kH - \bar{T}_{lc}\leq \frac{1}{2}-kH$. One  special case that comes to the surface is $\xi_3=\xi_2=1/2$, because the energy conservation is immediately visible. The potential energy of the chain elements in the beaker and those on the bottom are exactly the same at any given time and the carts balance each other out. Furthermore, the number of chain elements that are in motion stays the same. Consequently, it is expected that the existence of this case eliminates the possibility to describe the phenomenon with potential energy transfer. Due to the fact that $\bar{T}_0=\frac{1}{2} + \bar{T}_{lc} + kH/2$ is equal to $\bar{T}_0 = 1 - \frac{k}{q}  - k h$, the special case $\xi_3=\xi_2=1/2$ is obtained by connecting the purely kinematic parameters $h$, $H$ and $q$ with the parameter $k$ if $\bar{T}_{lc}=0$. \newline
Certainly, all other cases of interest can be addressed as well. The limit case $k = k_{\text{max}} = q/(1+q(h + H))$ for example is obtained for $\xi_3=1/(2kH)$ and $\bar{T}_{lc}=0$. It yields $\bar{T}_0=kH$ and as a result $\bar{T}_E=0$.
\begin{figure}[!htb]
\centering
\includegraphics[width=1\linewidth]{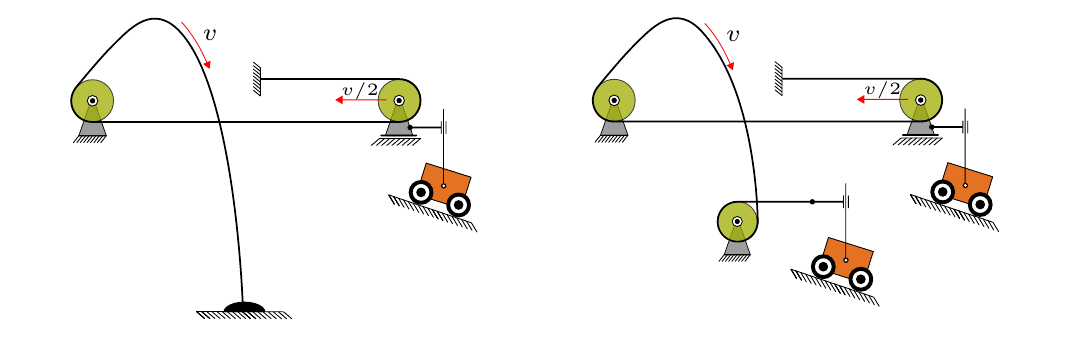}
\caption{Simplified pulley arrangements. Chain fountain with idealized beaker but real bottom (left). Further simplification of Fig.~\ref{fig:pulley_experiment}, but without subsequent deceleration of chain elements as more and more chain elements are in motion over time (right). Experiments with non-stationary fountains are straightforward to conduct by not guiding the lower part of the chain horizontally. Different steady-state chain velocities can be easily achieved by different lower cart masses.}
\label{fig:pulley_experiment_equiv_ground}
\end{figure}
Since the proposed substitute model supposedly does not precisely reflect all relationships of the chain fountain, as an approximation to the original experiment, one can omit the lower part of the arrangement and simply drop the chain on the floor as in the true experiment, see Fig.~\ref{fig:pulley_experiment_equiv_ground}. In any case, it is advantageous that the complex dynamic processes during collision of chain elements with the bottom can be studied separately to the pickup process. Experiments with non-stationary fountains could provide information about the material behaviour, e.g. it is certainly advantageous to observe the time-dependent system behaviour changes to quantify dissipative influences. As indicated in Fig.~\ref{fig:pulley_experiment_equiv_ground}, an arrangement can be specified in a simple manner in which the reaction force change is exactly known over time.